# Solution processed large area field effect transistors from dielectrophoreticly aligned arrays of carbon nanotubes

Paul Stokes, Eliot Silbar, Yashira M. Zayas[†] and Saiful I. Khondaker*

Nanoscience Technology Center & Department of Physics, University of Central Florida, 12424 Research Parkway, Orlando FL 32826, USA

Abstract

We demonstrate solution processable large area field effect transistors (FETs) from aligned arrays of carbon nanotubes (CNTs). Commercially available, surfactant free CNTs suspended in aqueous solution were aligned between source and drain electrodes using ac dielectrophoresis technique. After removing the metallic nanotubes using electrical breakdown, the devices displayed p-type behavior with on-off ratios up to $\sim 2 \times 10^4$. The measured field effect mobilities are as high as 123 cm$^2$/Vs, which is three orders of magnitude higher than typical solution processed organic FET devices.

* To whom correspondence should be addressed. E-mail: saiful@mail.ucf.edu

† Current Address: Chemical Engineering Department, University of Puerto Rico at Mayaguez, Mayaguez, P.R. 00681-9000

Solution processed electronic devices have attracted tremendous attention because of their ease of processablity, low cost of fabrication, and their ability to cover large areas. These devices may be useful for applications such as flexible displays, sensor sheets, radiofrequency (RFIDs) tags, and photovoltaics [1-2]. A significant amount of effort has been dedicated to improve device performance of solution processed organic field effect transistors (FETs). However, typical field effect mobilities for these devices are usually on the order of $\sim 0.1$ cm$^2$/Vs, and can very rarely reach $\sim 1.0$ cm$^2$/Vs [1-3]. In addition, the mobilities are highly sensitive to detailed fabrication parameters. For example, solution processable FETs made from the most commonly used polymer, regioregular poly(3-hexylthiophene) RR-P3HT, sensitively depends on the molecular weight, the dielectric-semiconductor interface, the solvent that it is spun from, surface treatments, post film formation treatment and annealing [4-7]. Furthermore, FET's made from polymers tend to degrade in air, adding another degree of difficulty to the procedure [8].

An alternative route to fabricate high quality solution processed FETs that can be superior to polymer based devices may be the use of carbon nanotubes (CNTs) dispersed in solution. FETs from individual CNTs have displayed exceptional electrical properties including subthreshold swings as low as 60 mV/dec and mobilities reaching 79,000 cm$^2$/Vs [9]. However, devices fabricated from arrays of CNTs can be advantageous over individual tube devices in certain cases, as they may provide more homogeniality from device to device and can cover large areas. In addition, devices fabricated with nanotube arrays contain hundreds of CNTs which can increase current outputs (up to hundreds of microamps). Large scale assembly of CNTs from solution can be achieved by several different techniques including chemical and biological patterning [10, 11], Langmuir-Blodgett assembly [12], bubble blown films [13], contact printing [14], ink-jet printing [15], spin coating assisted alignment [16], and evaporation driven self assembly [17]. All of these techniques create CNT networks where charge transport needs to occur through a large number of overlapping inter-nanotube contacts.

Recently, dielectrophoresis (DEP) has been used for the directed assembly of individual, bundles, or networks of CNTs [18-21]. However, high quality FETs from large area DEP assembled arrays has not been demonstrated. DEP assembled CNT-FET devices can be advantageous as every CNT connects between source and drain electrodes minimizing charge transport through CNT-CNT interconnects. Here we report on solution processed, large area high quality FETs from dielectrophoreticaly aligned arrays of CNTs. Commercially available, surfactant free CNT solution (suspended in DI water) [20] were assembled between source (S) and drain (D) electrodes patterned on a Si/SiO$_2$ substrate by applying an AC electric field. The highly doped Si substrate was used as a global back gate (G). After using an electrical breakdown technique to remove the metallic CNT pathways, the devices showed on-off ratios ($I_{on}/I_{off}$) up to ~ $2\times10^4$ with p-type FET behavior. The measured mobilities are as high as 123 cm$^2$/Vs which is three orders of magnitude higher than typical solution processed organic FETs. Our technique represents a simple and convenient way to fabricate high quality solution processable FET devices.

Highly doped Si wafers with 250 nm capped layer of SiO$_2$ were used as substrates. Source and drain electrodes with spacing $L$=5 μm and a width $W$=200 μm were defined using electron beam lithography (EBL), then electron beam evaporation of Cr (5 nm) and Pd (30 nm), followed by standard lift-off in acetone. The sample was then placed in oxygen plasma cleaner for 10 minutes to remove the unwanted organic residues on the surface.

Surfactant free, highly purified CNT's suspended in DI water were obtained from Brewer Science Inc. [22]. The obtained solution has a concentration of ~ 50 μg/ml. The solution is further diluted in DI water to obtain a concentration of ~1.0 μg/ml. The assembly of CNTs was carrier out in a probe station under ambient conditions. Figure 1a shows a schematic of the DEP assembly circuit. First, a 3 μL drop of the NT suspension was cast onto the electrode array. An AC voltage of 300 kHz, 5 Vp-p is applied between the source and gate electrode for 15 seconds. For high frequencies ($f$) of the AC voltage applied between source and gate, the impedance ($Z=1/j\omega C_{plate}$, $\omega = 2\pi f$) reduces considerably. Therefore the drain becomes capacitivly coupled to the gate electrode and obtains a similar potential as the gate electrode creating the necessary potential difference between the source and drain electrodes. Hence, the AC voltage creates a time averaged dielectrophoretic force between source and drain to align the CNTs. For an elongated object it is given by $F_{DEP} \propto \varepsilon_m \text{Re}[K_f] \nabla E_{RMS}^2$, $K_f = (\varepsilon_p^* - \varepsilon_m^*)/\varepsilon_m^*$, $\varepsilon_{p,m}^* = \varepsilon_{p,m} - i(\sigma_{p,m}/\omega)$ where $\varepsilon_p$ and $\varepsilon_m$ are the permittivity of the nanotube and solvent respectively, $K_f$ is the Claussius-Mossotti factor, $\sigma$ is the conductivity [18]. The induced dipole moment of the nanotube interacting with the strong electric field causes the nanotubes to move in a translational motion along the electric field gradient and align between the source and drain electrodes. As the CNTs assemble between source and drain electrodes, the parallel plate capacitance ($C_{plate}=\varepsilon A/t_{ox}$, $t_{ox}$ is the thickness and $\varepsilon=3.9\varepsilon_0$ is the dielectric constant of the SiO$_2$ layer) of the electrode/SiO$_2$/Si stack increases due to an increase of the effective area $A=WL$. This causes a decrease of the impedance of the drain/SiO$_2$/Si stack. This was evident in the observation of the output signal on the oscilloscope as it increased by 30-40% by the end of the assembly. Figure 1b shows an AFM image for a portion of a device after the assembly. The density of the aligned array is ~1 CNT/μm on average giving ~200 CNTs total in the channel. By varying the CNT density of the solution and the trapping time it is possible to tune the number of CNTs/μm in the array. A detailed study of this assembly will be presented elsewhere. The diameter of the CNTs, measured by AFM varies from 1.5 to 6.0 nm. A total of 16 devices were measured, of which half of the devices are measured as-assembled without further processing (bottom contacted device). The other half was measured following an additional EBL step for which 30 nm thick Pd was evaporated to form a top contact (top contacted device).

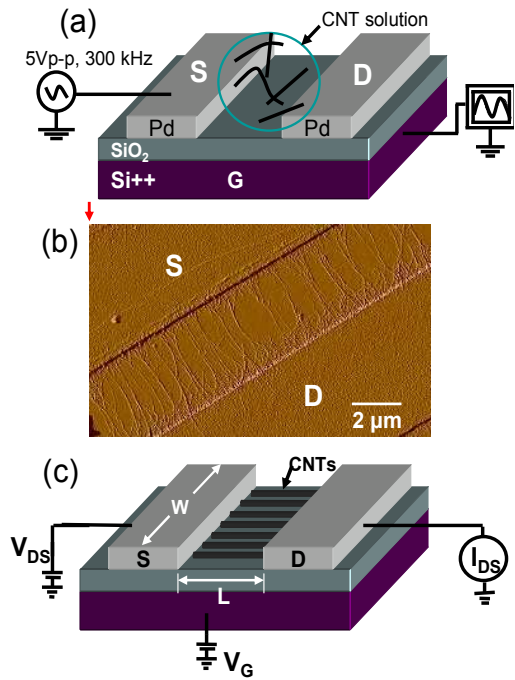

**Figure 1.** (color online) (a) Schematic of dielectrophoretic assembly. An AC voltage of 5 V at 300 kHz is applied to the source (S) electrode while the conducting Si substrate is monitored by an oscilloscope. (b) AFM image of a section of a device after assembly. (c) Cartoon for electronic transport measurements set up.

After the assembly the room temperature electronic transport measurements were carried out in a probe station using the Si substrate as a global back gate. Figure 1c shows a schematic of the electrical transport measurement which was performed by means of a Keithley 2400 sourcemeter, 6517A electrometer, and a current preamplifier interfaced with LabView. The initial two terminal resistance is typically in the range of 20-50 kΩ for bottom contacted devices and 2-5 kΩ for the top contacted devices. The mobility is calculated using the formula $\mu = (LA/WV_{DS}C)/(dI_D/dV_G)$. The capacitance $C$ of the CNT FET array device was approximated from

$$C = \frac{A \cdot D}{\left[C_Q^{-1} + \frac{1}{2\pi\varepsilon}\ln\left[\frac{\sinh(2\pi t_{ox} D)}{\pi R D}\right]\right]},$$

where $C_Q = 4 \times 10^{-10}$ F/m is the quantum capacitance, $R$ is the radius of the nanotubes, and $D$ is the linear density in CNTs per μm of the array [23]. According to this equation, the capacitance increases with increasing density of the nanotube and saturates to parallel plate capacitance value at high enough nanotube density. Here, we used $R = 1$ nm and $D = 1$ CNT/μm.

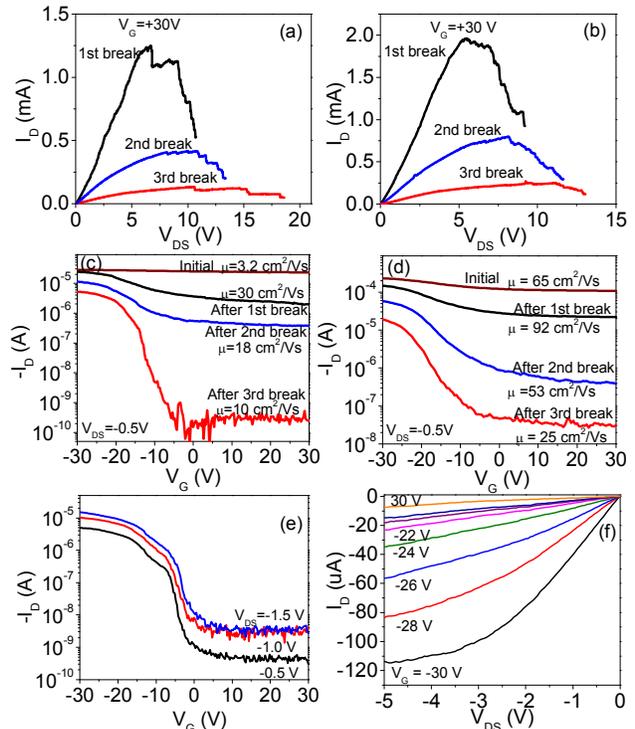

**Figure 2.** (color online) A representative plot of drain current ($I_D$) versus source-drain voltage ($V_{DS}$) for three sequential breakdowns (1$^{st}$, 2$^{nd}$, and 3$^{rd}$ break) for (a) a bottom contacted and (b) a top contacted device (c) $I_D$ versus back gate voltage $V_G$ at constant $V_{DS}$ of -0.5 V after each breakdown for the bottom contacted device. (d) $I_D$ vs. $V_G$ after each breakdown for top contacted device. (e) $I_D$ vs. $V_G$ after 3$^{rd}$ breakdown at different $V_{DS}$ of -0.5, -1.0, and -1.5 V for the same device as in 2c. The on-off ratio for this device is ~ $2\times 10^4$. (f) Output characteristics for the top contact device presented in 2d after 3$^{rd}$ breakdown.

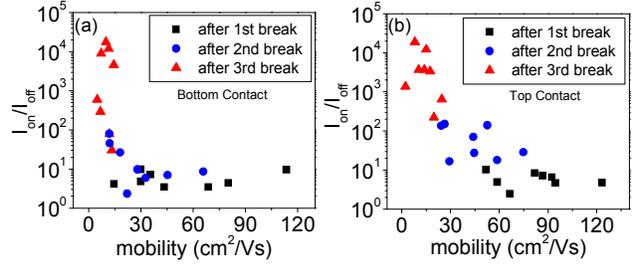

**Figure 3.** (color online) Plot of on-off ratios and corresponding mobility for all measured devices after each breakdown in (a) bottom contact and (b) top contact configuration.

The as-assembled aligned CNT array devices show semi-metallic behavior with on-off ratios, $I_{on}/I_{off} \sim 1.3$ to $3.0$ and average mobilities of $\mu \sim 5.5$ and $\sim 54$ cm$^2$/Vs for the bottom contact and top contact devices respectively. The low on-off ratio and modest mobility is due to the presence of large amount of metallic pathways in the array. Therefore to increase device performance, we performed an electrical breakdown procedure to controllably reduce the metallic pathways [24].

Figure 2a and 2b shows a representative plot of drain current ($I_D$) versus source-drain voltage ($V_{DS}$) for three sequential breakdowns (1$^{st}$, 2$^{nd}$, and 3$^{rd}$ break) for a bottom contacted and a top contacted device respectively. The back gate was held constant at $V_G = +30$ V to deplete the carriers in the p-type semiconducting CNTs while we ramped up $V_{DS}$ to eliminate the metallic CNTs. As $V_{DS}$ is ramped up, the CNTs start to breakdown and $I_D$ begins to fall. In order to obtain reproducible results, each breakdown is stopped when $I_D$ is about 50% of its peak value at which point $V_{DS}$ is swept back to zero. When the 3$^{rd}$ breakdown reaches ~50% of its peak value, $I_D$ can range from ~ 0.03 to 0.12 mA.

Figure 2c and 2d show $I_D$ versus $V_G$ characteristics for a typical bottom and top contact device respectively after each breakdown. The source-drain voltage is held constant at $V_{DS} = -0.5$ V. In figure 2c, the upper most curve is the initial sweep showing a mobility of 3.2 cm$^2$/Vs with very little on-off ratio (~1.1). After the first breakdown the field effect behavior of the device is enhanced - both the mobility and on-off ratios are increased to 30 cm$^2$/Vs and ~ 10 respectively due to a reduction of metallic pathways. After the second breakdown the mobility reduces a small amount to 18 cm$^2$/Vs and the on-off ratio increases to ~ 26. Finally, after the third breakdown the mobility is reduced to 10 cm$^2$/Vs, however the on-off ratio increases a few orders of magnitude to ~ $2 \times 10^4$. Figure 2d shows similar behavior for the top contacted device with $\mu$ = 65, 92, 53, and 25 cm$^2$/Vs and on-off ratios of 2.1, 6.6, 14, and 650 for the initial sample and then after the 1$^{st}$, 2$^{nd}$, and 3$^{rd}$ breakdown respectively. We find that the top contacted devices show higher mobilities which is most likely due to the reduced contact resistance. Figure 2e is a plot of $I_D$ versus $V_G$ at different $V_{DS}$ after the third breakdown for the same device shown in figure 2c. Figure 2f shows the detailed output characteristics, $I_D$ versus $V_{DS}$ at different $V_G$ recorded for the sample presented in figure 2d after the third breakdown.

Figure 3a and 3b shows the on-off ratio and corresponding mobility value for all of the bottom (figure 3a) and top contact (figure 3b) devices after each breakdown. In figure 3a the on-off ratio remains fairly constant and then increases more rapidly after the 3$^{rd}$ breakdown with median on-off ratios after each breakdown of 4.6, 9.2, and $2.6 \times 10^3$. Figure 3b shows a more steadily increase in on-off ratio after each breakdown with median values of 5.8, 50, $3.5 \times 10^3$.

For the bottom contacted devices, the median mobilities are 50, 27, and 9.1 cm$^2$/Vs after first, second, and third breakdown respectively. Top contacted devices yield median mobility values of 77, 41, and 15 cm$^2$/Vs after the three breakdowns respectively. We found that, the top contacted devices are more controllable and show better device to device reproducibly after each breakdown. This is most likely due to the better contact resistance from the top contact. The highest mobility obtained from all the devices is 123 cm$^2$/Vs. The mobility values reported here are up to three orders of magnitude higher than typical FET devices made from solution processed polymers [2].

In conclusion we have demonstrated solution processable large area field effect transistors (FETs) from aligned arrays of CNTs. The CNTs were aligned from a commercially available, surfactant free CNTs suspended in aqueous solution using AC dielectrophoresis. After reducing the metallic pathways using electrical breakdown, the devices displayed on-off ratios up to $\sim 2\times 10^4$. The devices showed p-type FET behavior with mobilities up to three orders of magnitude higher than typical solution processed organic FET devices. The ease of processing for the dielectrophoreticly assembled devices presented here offers an alternative to solution processed polymer FET devices.

This work is partially supported by US National Science Foundation under grants ECCS-0748091 (CAREER) and NSF REU Site: EEC-0453436.